\documentclass[aps,prd,nopacs,floatfix,notitlepage,superscriptaddress,nofootinbib,twocolumn,a4paper,longbibliography]{revtex4-1}
\usepackage{amsfonts,amsmath,units,wasysym,epsfig,graphicx,verbatim,color,subfigure,graphicx,bm,mathrsfs,lipsum,hyperref,cleveref}
\usepackage[utf8]{inputenc}
\usepackage{booktabs}
\usepackage[normalem]{ulem}  
\usepackage{xcolor}
\usepackage{enumerate}

\newcommand{\BITS}{\affiliation{Department of Physics, BITS Pilani, Hyderabad Campus, India}}
\newcommand{\PU}{\affiliation{Department of Physics, Panjab University, Chandigarh, India}}
\newcommand{\USAL}{\affiliation{Departamento de F\'isica Fundamental and IUFFyM, Universidad de Salamanca, Plaza de la Merced S/N, E-37008 Salamanca, Spain}}
\newcommand{\Uliege}{\affiliation{Space Sciences, Technologies and Astrophysics Research (STAR) Institute, Universit\'e de Li\`ege, B\^at. B5a, 4000 Li\`ege, Belgium}}

\begin{document}

\title{Systematic study of scalar, vector, and mixed density dependencies in relativistic mean-field descriptions of hyperonic matter in neutron stars}

\author{Aprajita Shrivastava}
\email{p20230071@hyderabad.bits-pilani.ac.in}
\BITS

\author{Prasanta Char}
\email{prasanta.char@usal.es}
\USAL \Uliege

\author{Sakshi Gautam}
\email{sakshigautam@pu.ac.in}
\PU

\author{Sarmistha Banik}
\email{sarmistha.banik@hyderabad.bits-pilani.ac.in}
\BITS

\begin{abstract}
We investigate the equation of state (EOS) of hyperonic neutron star(NS) matter within a density-dependent relativistic mean-field (DDRMF) framework. The effects of scalar, vector, and mixed density dependencies in meson–baryon couplings are systematically examined along with alternative forms of the $\rho$-meson coupling. Several meson–nucleon parameter sets are explored here for the first time for the NSs and compared with the standard DD2 EOS. Most new parameterizations produce stiffer EOSs, leading to NSs with larger radii and higher tidal deformabilities. However, the inclusion of $\Lambda$ hyperons softens these EOSs, and  the resulting maximum masses still satisfy the 2$M_{\odot}$ limits and  agree with NICER measurements.  These results highlight the importance of exploring alternative density dependencies in constraining dense matter through multi-messenger observations.
\end{abstract}

\maketitle

\section{Introduction}
The study of nuclear matter under extreme conditions of density and isospin asymmetry is fundamental to our understanding of a wide range of astrophysical phenomena, particularly the structure and evolution of neutron stars (NSs) \cite{Glendenning:1997wn}. Among the theoretical frameworks developed to describe the properties of dense nuclear matter, Relativistic Mean Field (RMF) models have emerged as a powerful and consistent approach grounded in quantum field theory \cite{Walecka:1974qa,Boguta:1977xi,Serot:1984ey}. These models describe nucleons interacting through the exchange of mesons, and they naturally incorporate key features such as Lorentz covariance and the saturation properties of nuclear matter \cite{Dutra:2014qga,Sun:2023xkg}.

Traditional RMF models, with constant meson-nucleon coupling constants, have been successful in describing finite nuclei and symmetric nuclear matter \cite{Dutra:2014qga}.
However, their limitations become evident when extrapolated to high-density or extreme isospin asymmetry regimes, such as those found in the core of neutron stars. In response, 
Density-Dependent RMF (DDRMF) models have been developed \cite{Fuchs:1995as, Keil:1999hk,Hofmann:2000vz}, where the meson-nucleon couplings are explicitly dependent on the baryon density. By introducing density dependence in the meson-nucleon coupling constants, particularly in the vector (baryon) density, it becomes possible to more accurately reproduce empirical nuclear matter properties, such as the saturation point, compressibility, and symmetry energy, as well as to better constrain the equation of state (EOS) at supra-saturation densities \cite{Typel:1999yq}. 
This approach is based on the fact that effective interactions between nucleons are modified by the surrounding medium, especially under varying density conditions.  Moreover, vector density dependence is typically employed because it maintains Lorentz covariance and leads to thermodynamically consistent equations of motion. 

Fuchs \textit{et al.}~\cite{Fuchs:1995as} also introduced a scalar density dependence of the meson--baryon couplings. Building on this, Typel~\cite{Typel2018} investigated several formulations of the DDRMF models, incorporating scalar-, vector-, and mixed-density dependencies. The scalar density, which characterizes the local distribution of mass--energy in the nuclear medium, offers a distinct sensitivity to in-medium modifications, particularly at higher densities. It is observed that while all parameterizations (vector-, scalar-, or mixed-density dependent) yield comparable results at subsaturation densities, significant variations appear in the nuclear matter parameters and the EOS at higher densities. Scalar density-dependent models generally predict a stiffer symmetry energy ($E_{\text{sym}}$) and larger slope parameters ($L$), whereas vector-dependent models tend to produce softer trends. In particular, the incompressibility ($K$), skewness ($Q$), and symmetry incompressibility ($K_{\text{sym}}$) exhibit strong correlations with the functional form and choice of density dependence. Although these studies were primarily aimed at and limited to exploring uniform hadronic matter, they nevertheless provided valuable insights into the properties of dense matter in NS interiors. Later, the DDRMF theory was further applied to NS matter~\cite{Hofmann:2000mc}, incorporating conditions of $\beta$-equilibrium and charge neutrality to describe matter at supranuclear densities. This framework naturally paved the way for extensions of the RMF approach to include additional degrees of freedom in the dense cores of NSs\cite{Banik2002}.
\par
RMF models are also extended beyond nucleonic degrees of freedom, especially in the inner core of NS \cite{Schaffner:1995th}. For example, hyperons (strange baryons) are expected to appear in NS cores at densities exceeding 2-3 times the nuclear saturation density due to a very high baryon chemical potential in such dense environments. The appearance of hyperons is thus energetically favored, leading to a softening of the EOS and a decrease in the predicted maximum mass \cite{Banik2014}. However, the observation of massive NS ($\sim 2M_\odot$), such as PSR J0348+0432 \cite{Antoniadis:2013pzd} and PSR J0740+6620 \cite{Fonseca:2021wxt}, poses a challenge to this prediction, giving rise to the ``hyperon puzzle". Ongoing efforts continue to focus on improving our understanding of hyperon–nucleon and hyperon–hyperon interactions, which remain poorly constrained because of the scarcity of experimental data \cite{Banik2014, Malik2021}. Therefore, studying hyperons within the RMF framework helps in understanding dense matter behavior, testing the limits of theoretical models, and reconciling theory with astrophysical observations. 
Over the past several years, many different types of covariant density functionals have been explored using Bayesian techniques incorporating multiple theoretical, experimental and observational constraints \cite{Traversi:2020aaa,Malik:2022zol,Malik:2022jqc,Beznogov:2022rri,Providencia:2023rxc,Raduta:2024awt,Parmar:2024qff,Li:2025oxi,Passarella:2025zqb,Li:2025vhk}.

Motivated by the developments in DDRMF models in nuclear matter, the vector density dependence in NS matter has been thoroughly investigated by several groups; however, the scalar density dependence has not yet been systematically explored in this context. Recently, Ref.~\cite{Li2023} examined a mixed-density dependent scenario, though limited to purely nucleonic matter. 
The present work complements these earlier studies by extending the formalism to include hyperons and by examining NS properties under different density-dependent coupling schemes—namely, scalar, vector, and mixed—considering both purely nucleonic and hyperonic compositions in the stellar core.

The structure of the paper is as follows. In section \ref{formalism}, we revisit the formalism to calculate the NS EOS and structure. We discuss different types of density dependencies in \ref{dd}, the calculation of the EOS with different coupling choices in \ref{param}, and NS structure calculation in \ref{tov}. We report our results and discuss their implications in section \ref{results}. Finally, we summarize and conclude in section \ref{summary}.

\section{Formalism}
\label{formalism}

The traditional starting point to derive a relativistic mean field model
is a Lagrangian density $\mathcal{L}$
with baryons (neutron, protons and $\Lambda$ hyperons) and mesons (\(\sigma\), \(\omega\),  \(\rho\) and additional \(\phi\) for the hyperon-interactions exclusively) as degrees of freedom. Here, we follow mostly
the notation as in Ref. \cite{Typel2018}:

\begin{eqnarray}
\label{eq:L}
\mathcal{L_B} &=& \sum_{B=N, \Lambda}  \bar{\psi}_B \big( i\gamma_\mu \partial^\mu - m_B + g_{\sigma B} \sigma 
- g_{\omega B}  \gamma_\mu \omega^\mu  \notag \\
&-&  g_{\rho B}  \gamma_\mu \boldsymbol{\tau} \cdot \boldsymbol{\rho}^\mu -g_{\phi B} \gamma_{\mu} \phi^{\mu} \big) \psi_B  
+ \frac{1}{2} (\partial_\mu \sigma \partial^\mu \sigma - m_\sigma^2 \sigma^2)\notag\\
&-& \frac{1}{4} \omega_{\mu\nu} \omega^{\mu\nu} 
+   \frac{1}{2} m_\omega^2 \omega_\mu \omega^\mu 
- \frac{1}{4} \rho_{\mu\nu} \cdot \rho^{\mu\nu} 
+ \frac{1}{2} m_\rho^2 \boldsymbol{\rho}_\mu \cdot \boldsymbol{\rho}^\mu\notag\\
&-& \frac{1}{4} \phi_{\mu\nu}\phi^{\mu\nu} +  \frac{1}{2} m_\phi^2 \phi_\mu \phi^\mu .
\end{eqnarray} 
Here, \(\bar \psi_B =\Psi_{B}^{\dagger}\gamma^{0}\) is the Dirac field of the baryons with rest mass $m_B$;  their coupling to the meson fields are density dependent and are denoted 
by isoscalar  \(g_{\sigma B}\), vector \(g_{\omega B}\), \(g_{\phi \Lambda}\) and isovector  \(g_{\rho B}\). The vector meson interactions are  $\mathbf {R}^{\mu\nu}  =  \partial^{\mu}  \mathbf{R}^{\nu} - \partial^{\nu}\mathbf{R}^{\mu}$ where $\mathbf{R}=\omega,  \rho$ and $\phi$. \(\gamma_\mu\)'s  are the standard relativistic matrices  whereas $\vec{\tau}$ are the isospin matrices.   $\mathcal{L}_l = \sum_l \bar \psi_l (i \gamma_{\mu} \partial ^{\mu} -m_l)\psi_l$ is the Lagrangian for non-interacting leptons ($l= \mu$, $e^-$).

From the Lagrangian density the field equations
of all degrees of freedom are found using the Euler-Lagrange
equations. Applying the  mean-field approximation
and exploiting the symmetries of a stationary system, the field equations assume a simple form.  Note that only the time-like components of the vector fields ($\omega_0$, $\phi_0$)  and the third
isospin component of the isovector field $\rho_{03}$  survive in a uniform and static matter. The field equations for different mesons are,
\begin{eqnarray}
m_{\sigma}^2 \sigma &=& \sum_B g_{\sigma B}n^s_B, \notag\\
m_{\omega}^2 \omega_0 &=& \sum_B g_{\omega B}n^v_B, \notag\\
m_{\rho}^2 \rho_{03} &=&  \sum_B g_{\rho B} \tau_{3B} n^v_B,\notag\\ 
m_{\phi}^2 \phi_0 &=& g_{\phi \Lambda} n^v_{\Lambda}.
\label{Eq:meson}
\end{eqnarray}
The couplings  $g_{jB}$'s are density dependent, where $j=\sigma$, $\omega$, $\rho$ or $\phi$ and 
\begin{eqnarray}
   n^{s}_B &=& <\bar \psi_B \psi_B>\notag\\
   &=&\frac{m_B^*}{2\pi^2}\left[ 
    k_{F_B} \sqrt{k_{F_B}^2 + m_B^{\ast2}} - m_B^{\ast2} \ln \frac{k_{F_B} + \sqrt{k_{F_B}^2 + m_B^{\ast2}}}{m_B^\ast}
    \right], \notag \\
    n^{v}_B&=&<\psi_B ^\dagger \psi_B> = \frac{k_{F_B}^3}{3 \pi^2},
    \label{eq:scalarvector}
\end{eqnarray}
are the scalar and vector densities, respectively.
Here $k_{F_B}$ is the Fermi momentum of the baryon B.
The baryon field equation turns out to be
\begin{equation}
    [\gamma_{\mu} (i \partial ^{\mu} -  V_B) - m_B^\ast]\psi_B=0.
    \label{Eq:Dirac}
\end{equation}
Here, the effective mass is $m_{B}^{\ast} =  m_{B} - S_B$, where $S_B(V_B)$ are the scalar (vector) potential, respectively and are expressed as
\begin{equation}
    S_B = g_{\sigma B} \sigma + \Sigma^r_B ,    \label{Eq:Sb} 
\end{equation}
\begin{equation}
    V_B = g_{\omega B} \omega_{0} + g_{\rho B} \rho_{03} +  g_{\phi B} \phi_0 - \Sigma^r_B.
    \label{Eq:Vb}
\end{equation}
In the expression of $V_B$ of Eq. \ref{Eq:Vb}, $\Sigma^r_B$ is summed over all the baryons. 
The rearrangement term $\Sigma^r_B$ for a baryon ($B = N, \Lambda$) in Eqs: \ref{Eq:Sb} and \ref{Eq:Vb}, are  scalar  and vector density dependent respectively, and are given by:

\begin{eqnarray}
\Sigma^r_B &=&\Sigma^r_S -\Sigma^r_V \nonumber \\
&=&\left. \frac{\partial g_{\sigma B}}{\partial n} \right|_{n'} n_{\sigma}\sigma - \left. \frac{\partial g_{\omega B}}{\partial n}\right|_{n'} n_{\omega}\omega_0- \left. \frac{\partial g_{\rho B}}{\partial {n}} \right|_{n'} n_\rho \rho_{03} \notag\\
 &-& \left. \frac{\partial g_{\phi B}} {\partial n} \right|_{{n'}} n_\phi \phi_0 
 \label{Eq:Sr}
 \end{eqnarray}
where $\Sigma^r_S$ and $\Sigma^r_V$ denote the scalar and vector rearrangement contributions, respectively.
Here,  $n_\sigma = \sum_B n^s_B$, $n_\omega = \sum_B n^v_B$, $n_\rho = (n^v_p - n^v_n)$, and $n_\phi=n^v_\Lambda$ is the vector number density exclusive to the $\Lambda$ hyperon. 
The couplings are differentiated with respect to the density ($n=n^s$ or $n^v$) in Eq. \ref{Eq:Sr}, at a fixed $n'$ where $n' = n^s$ when $n = n^v$and vice versa, depending on whether the couplings are chosen to depend on the vector or scalar density.

The meson field Eqns: \ref{Eq:meson} are solved simultaneously along with constraints of charge neutrality, total baryon number conservation and $\beta$-equilibrium.
The chemical potential of the baryon B is given by
\begin{equation}
\mu_B  =  \sqrt{k_{F{_B}}^2 +m_B^{\ast 2}} + V_B
\label{Eq:mu}
\end{equation}
The $\Lambda$ hyperons populates the core when $\mu_n=\mu_\Lambda$ condition is satisfied.

We consider three cases of density dependence for the couplings:
\begin{enumerate}[(I)]
    \item Scalar Density Dependent (SDD)
    \item Vector Density Dependent (VDD)
    \item Mixed Density Dependent (MDD)
\end{enumerate}

We will describe them briefly in the next section.

\subsection{Types of density dependent couplings}
\label{dd}

All the meson-baryon couplings depend on the number density in the form  $g_{j B}(x)$, where $x$ is the scaling factor defined as $x=n/n_{ref}$, and $j=\sigma$, $\omega$, $\rho$ or $\phi$ . Here $n_{ref}$ is saturation number density. 

In SDD (VDD) case, all the couplings depend on scalar (vector) number density, i.e. $x =\frac {n^s}{ n_{ref}^s} (\frac {n^v}{n_{ref}^v})$. \\
In the case of MDD, $\sigma$ meson-baryon coupling depends on scalar number density, 
where the rest of the couplings depend on vector number density. 
The values of $n_{ref}$'s are taken from the table: A1 from Ref. \cite{Typel2018}.

The density-dependent couplings are written in the form $g_{j }= g_{j }^0 f_j(x)$   where  $g_j^0=g_j(n_{ref})$ and $f_j(x)$ is functional form of couplings that we discuss next.

\subsection{Parametrization of Couplings}
\label{param}

We start discussing the parametrization for meson-nucleon couplings first.
Following Ref. \cite{Typel2018}, the functional forms of $f(x)$ are taken  i) rational (R):
\begin{equation}
    f_j(x)=a_j \frac {1 +b_j (x +d_j)^2} {1 +c_j (x +d_j)^2},
    \label{Eq:funcR}
\end{equation}
and ii) exponential (E):
\begin{equation}
f_j(x)= \exp{[-a_j (x - 1)]}.
\label{Eq:funcE}
\end{equation}
For $\sigma$, $\omega$ and $\phi$ mesons we consider R dependence, while for $\rho$ mesons we choose either R or E forms.
Consequently, we denote our models as SZE, SZR, MZE, MPE, MZR, and VZR, following the terminology of \cite{Typel2018}. Here, S, M, and V refer to scalar, mixed, and vector density models, respectively. The Z or P designation depends on the value of the parameter $d$ which can be either positive (P) or zero (Z), ie. $d\geq0$ in Eq. \ref{Eq:funcR}. Note that $d < 0$ corresponded to collapsing effective mass as reported in Ref. \cite{Typel2018} and thus such parameterizations are avoided. Explicit values of the parameters $b$, $c$, $d$, 
and masses of mesons are taken from A1-A3 tables of Ref \cite{Typel2018}. They are obtained by  fitting of DDRMF models with different density dependencies of the couplings to observable properties of nuclei such as binding energies per nucleon, charge radii, diffraction radii, surface thicknesses and spin-orbit splittings. The  parameter $a_j$ for the rational form in Eq. \ref{Eq:funcR} is constrained by the condition $f_j(1) = 1$ such that 
$a_j=  \frac {1 +c_j (1 +d_j)^2} {1 +b_j (1 +d_j)^2}$.

Next, we determine the hyperon-meson couplings. 
The vector coupling constants for hyperons are determined from the SU(6)
symmetry \cite{Schaffner:1995th} as,
\begin{eqnarray}
\frac{1}{2}g_{\omega \Lambda} =
\frac{1}{3} g_{\omega N},\nonumber\\
g_{\rho \Lambda} = 0, \nonumber\\
2 g_{\phi \Lambda} =
-\frac{2\sqrt{2}}{3} g_{\omega N} ~.
\end{eqnarray}
These vector meson coupling are plugged into the expressions of the potential $U_{\Lambda}^N$ at saturated nuclear matter density, which are related to the meson fields by: \\
For SDD case:
\begin{equation}
U_{\Lambda}^N(n_{ref}) = - g_{\sigma \Lambda} {\sigma} + g_{\omega \Lambda} {\omega_0}.
\end{equation}
For  MDD case:
\begin{equation}
U_{\Lambda}^N(n_{ref}) = - g_{\sigma \Lambda} {\sigma} + g_{\omega \Lambda} {\omega_0} + \omega_0 n_w \frac{\partial g_{\omega N}}{\partial n^v}.
\end{equation}
For VDD case:
\begin{equation}
U_{\Lambda}^N(n_{ref}) = - g_{\sigma \Lambda} {\sigma} + g_{\omega \Lambda} {\omega_0} - \sigma n_\sigma \frac{\partial g_{\sigma N}}{\partial n^v} + \omega_0 n_w \frac{\partial g_{\omega N}}{\partial n^v} .
\end{equation}
From the above equations, we can find the scalar meson couplings to hyperons $g_{\sigma\Lambda}$ for $U_{\Lambda}^N$ =-30MeV \cite{millener}. 
For different models, $R_{\sigma\Lambda}=\frac{g_{\sigma\Lambda}}{g_{\sigma N}}$ is calculated and the values obtained are listed in Table \ref{tab:scaling}.
\begin{table}
\begin{ruledtabular}
\begin{tabular}{lcccccc}
\textrm{Models}&
\textrm{SZE}&
\textrm{SZR}&
\textrm{MZE}&
\textrm{MPE}&
\textrm{MZR}&
\textrm{VZR}\\
\colrule
$R_{\sigma\Lambda}$ & 0.61 & 0.61 & 0.52 & 0.56 & 0.52 & 0.62\\
\end{tabular}
\end{ruledtabular}
\caption{Scaling constant for $\sigma$ coupling for $\Lambda$ hyperons for different models.}
\label{tab:scaling}
\end{table}

The  energy density and pressure are related
to the energy-momentum tensor $\Gamma_{\mu \nu}$ through  $\epsilon =< \Gamma_{00} >$ and $P =< \Gamma_{\mu \mu} >$. 
$\Gamma_{\mu \nu}$ on the other hand is calculated from Eq. \ref{eq:L}
\begin{equation}
\Gamma_{\mu\nu} = -g_{\mu\nu} \mathcal{L} + \frac{\partial \mathcal{L}}{\partial (\partial^\mu \psi)} \partial_\nu \psi.
\end{equation}
In SDD model the general expressions for the energy density and pressure for the hadrons are:
\begin{eqnarray}
    \varepsilon &=&\sum_{B=N, \Lambda} \Big[ \varepsilon_B^{\text{kin}} +\sum_{j=\sigma, \omega, \rho, \phi}   \frac{1}{2} {m_j^2} j^2
 +  n_{\sigma}^2\sigma \frac{\partial g_{\sigma B}}{\partial n^s} \nonumber\\
 &&- \sum_{j=\omega, \rho, \phi}  n_{\sigma} j n_j \frac{\partial g_{jB}}{\partial n^s}  \Big]
 \label{Eq:Es}
    \end{eqnarray}
\begin{eqnarray}
P &=& \sum_{B = N, \Lambda} \Big[P^{\text{kin}}_B -\frac{1}{2} m_\sigma^2 \sigma^2-n^2_{\sigma} \sigma \frac{\partial g_{\sigma B}}{\partial n^s}\nonumber\\
&&+ \sum_{j=\omega, \rho, \phi} \left(\frac{1}{2} m_j^2 j^2 +   n_{\sigma} j n_j \frac{\partial g_{jB}}{\partial n^s} \right) \Big]
\label{Eq:Ps}
\end{eqnarray}
The kinetic terms in Eqs: \ref{Eq:Es} \& \ref{Eq:Ps} are, $\varepsilon_B^{\text{kin}} = \frac{1}{4} \left[ 3\mu_B^\ast n_B^{v} + m_B^\ast n_B^{s} \right]$,  $P_B^{\text{kin}} = \frac{1}{4} \left[ \mu_B^\ast n_B^{v} - m_B^\ast n_B^{s} \right]$ where the effective chemical potential  $\mu_B^\ast = \sqrt{k_{F{_B}}^2 +m_B^{\ast 2}}$.

For the MDD models, the respective expressions are:
\begin{equation}
    \varepsilon =\sum_{B=N, \Lambda} \left[ \varepsilon_B^{\text{kin}} + \sum_{j=\sigma, \omega, \rho, \phi}  \frac{1}{2} {m_j^2} j^2 
+  n^2_\sigma \sigma  \frac{\partial g_{\sigma B}}{\partial n^s}  \right] 
 \label{Eq:Em}
    \end{equation}

\begin{eqnarray}
P &=& \sum_{B = N, \Lambda} \Big[P^{\text{kin}}_B -  \frac{1}{2} m_\sigma^2 \sigma^2 - n^2_\sigma  \sigma \frac{\partial g_{\sigma B}}{\partial n^s} \nonumber\\&&+\sum_{j=\omega, \rho, \phi}  \left(\frac{1}{2} m_j^2 j^2 +
n_\omega j n_j \frac{\partial g_{jB}}{\partial n^v}\right) \Big]
\label{Eq:Pm}
\end{eqnarray}
Finally, in vector model, the $\varepsilon$ is same as Eq.\ref{Eq:Em} without the last term, while pressure is given by
\begin{eqnarray}
P &=&\sum_{B = N, \Lambda} \Big[P^{\text{kin}}_B -\frac{1}{2} m_\sigma^2 \sigma^2 -n_\omega \sigma n_\sigma \frac{\partial g_{\sigma B}}{\partial n^v} \nonumber\\
&&+\sum_{j=\omega, \rho, \phi} \left( \frac{1}{2} m_j^2 j^2 + n_\omega j n_j \frac{\partial g_{jB}}{\partial n^v} \right) \Big].
\label{Eq:Pv}
\end{eqnarray}
Leptons are considered non-interacting, and their contributions to energy density  and pressure are given by:
\begin{eqnarray}
\varepsilon_{\ell} = \sum_{\ell = e^-,\mu^-} \frac{1}{\pi^2} \int_0^{k_{F_\ell}} \left( k^2 + m_\ell^2 \right)^{1/2} k^2 \, dk\notag\\
P_{\ell} = \frac{1}{3} \sum_{\ell = e^-,\mu^-}\frac{1}{\pi^2} \int_0^{k_{F_\ell}} \frac{k^4 \, dk}{\left( k^2 + m_\ell^2 \right)^{1/2}}.
\end{eqnarray}
In this work, we employ the general-purpose crust EOS from density-dependent unified models~\cite{Hempel_2010, Banik_hempel_apj_2014}.
\subsection{NS Structure}
\label{tov}

The  tabulated EOS, obtained in the previous Section are used to calculate the observable parameters i.e. mass (M), radius(R)  by solving the Tolman–Oppenheimer–Volkoff (TOV) equations\cite{Oppenheimer,Tolman}:
\begin{eqnarray}
    \frac{dP(r)}{dr} &=& -\frac{(\mathcal{\varepsilon}(r) + P(r))(M(r) + 4\pi r^3 P(r))}{r^2 (1 - 2M(r)/r)}\\ \nonumber
     \frac{dM(r)}{dr} &=& 4\pi r^2 \mathcal{\varepsilon}(r).
     \label{Eq:TOV}
 \end{eqnarray}
The radius R is determined by the boundary condition i.e. pressure P = 0 on the surface.
When a spherical NS experiences a gravitational tidal field $E_{ij}$, it develops a quadrupole moment $Q_{ij}$ in response. To linear order, these quantities are related through the tidal deformability parameter $\lambda$ as: 
\begin{equation}
    \lambda = -\frac{Q_{ij}}{E_{ij}}.
\end{equation}
The tidal deformability can be expressed in terms of the star's radius R and the dimensionless Love number $k_2$ \cite{Hinderer}:
\begin{equation}
    \lambda = \frac{2}{3}{k_2 R^5}.
\end{equation}
The dimensionless tidal deformablity $\Lambda_{TD}$ is then defined as,
\begin{equation}
    \Lambda_{TD} = \frac{2}{3}{k_2C^{-5}}.
\end{equation}
Here, $C=M/R$ is the compactness of star of mass M and radius R.

\section{Results}
\label{results}

We report our results calculated using different coupling parameter sets as discussed above corresponding to scalar, vector, and mixed of both scalar and vector density dependence. Additionally, we have two different variations for the $\rho$-meson coupling functional. For consistency, we adopt  the nomenclature of the parameter sets defined in Ref. \cite{Typel2018}. The EOS of cold matter at $\beta$-equilibrium is first constructed for nucleonic matter consisting of neutrons, protons, electrons, and muons. Subsequently, we extend the calculation to include  $\Lambda$ hyperons within the SU(6) symmetry. At higher densities, the neutron chemical potential increases and eventually exceeds the rest mass of the $\Lambda$ hyperon. Once this threshold is reached, $\Lambda$ hyperons begin to populate the system rapidly, thereby lowering the ground-state energy in accordance with the Pauli exclusion principle. 

\begin{figure*}
    \includegraphics[scale=0.50]{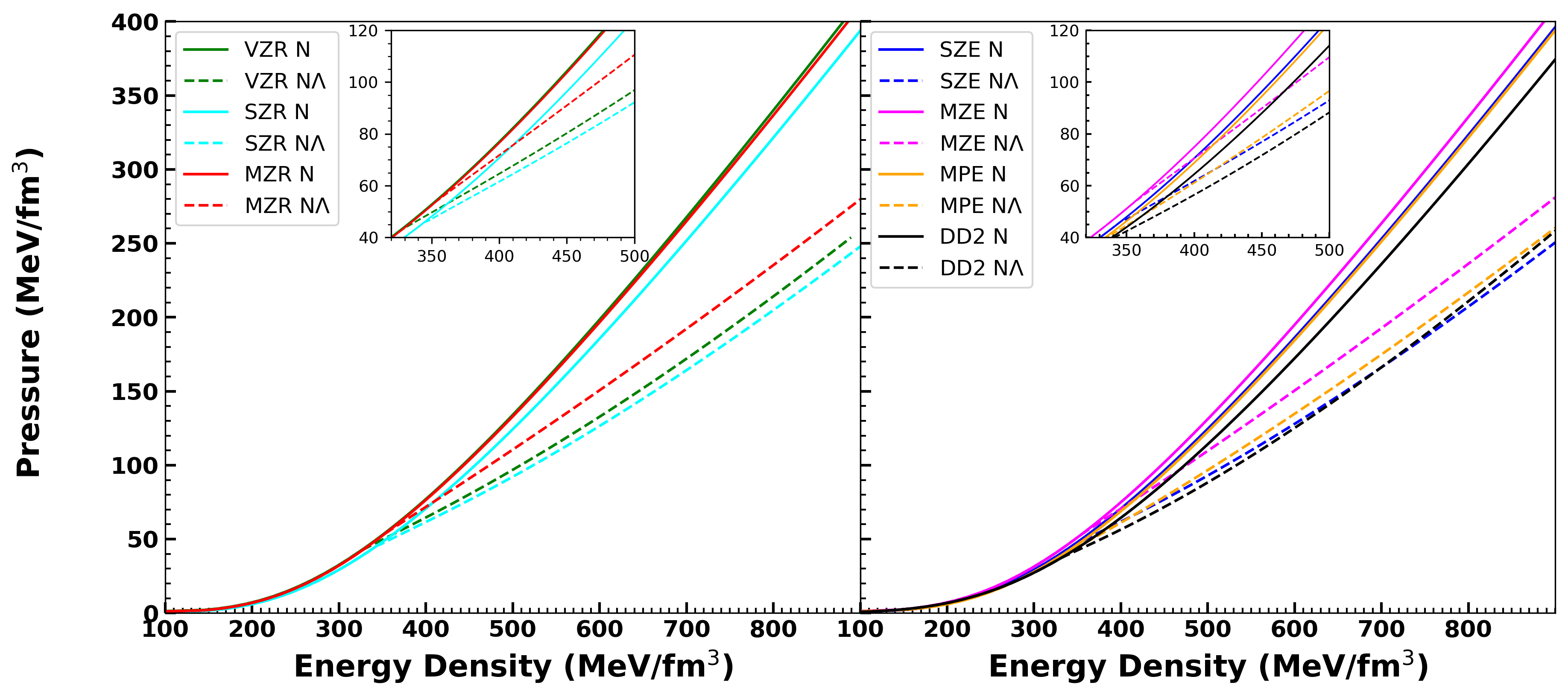}
    \caption{Equation of State (EOS) for various models. Left panel VZR, SZR \& MZR sets. Right panel SZE, MZE, MPE
 and DD2 sets. Solid lines and dashed lines represent nucleonic (N) and hyperonic (N$\Lambda$) models, respectively. The colour scheme is indicated in the figure. Inset highlights the EOS behaviour at lower densities, where SZR (DD2-N) is softer than MZR-N$\Lambda$ (MZE-N$\Lambda$) EOS.}
    \label{fig:EOS}
\end{figure*}

In Fig. \ref{fig:EOS}, we have shown the EOSs for different parameter sets; nucleonic with solid lines and the corresponding hyperonic ones with dashed lines. To decongest the plots, we show the EOS plots  in two panels. We show the sets with rational functional for the $\rho$ coupling on the left panel and the exponential functional for the same on the right panel. We have also added the EOS for the standard DD2 model  and its hyperon counterpart for reference \cite{Typel:2009sy, Banik2014, Char2014}. We have maintained this style in all the figures throughout the article.

From Fig.  \ref{fig:EOS}, we observe the general softening of the EOS following the onset of hyperons. In the left panel, we see that  VZR is stiffer than MZR and SZR for nucleonic case. However, after the inclusion of $\Lambda$s, SZR becomes slightly softer than VZR while MZR with $\Lambda$s remains quite stiff. In the right panel, the MZE is the stiffest among the nucleonic EOSs, while the SZE and MPE nearly overlap. After adding $\Lambda$s, MZE continues to remain stiffest, but the overlap breaks between SZE and MPE at high density as SZE becomes softer of the two. DD2 nucleonic model appears as considerably softest of them all. Consequently, DD2 with $\Lambda$s is also found to be the softest of all the hyperon EOSs considered here. Interestingly,  until 410 (465) MeV/fm$^{-3}$, the SZR (DD2-N) EOS is softer than MZR-N$\Lambda$ (MZE-N$\Lambda$) EOS as highlighted in inset in left (right) panels. All sets of EoS are checked for thermodynamic consistency using Gibbs-Duhem Relation: $\varepsilon + P = \sum_B n_B \mu_B$.

\begin{figure*}
     \includegraphics[scale = 0.50]{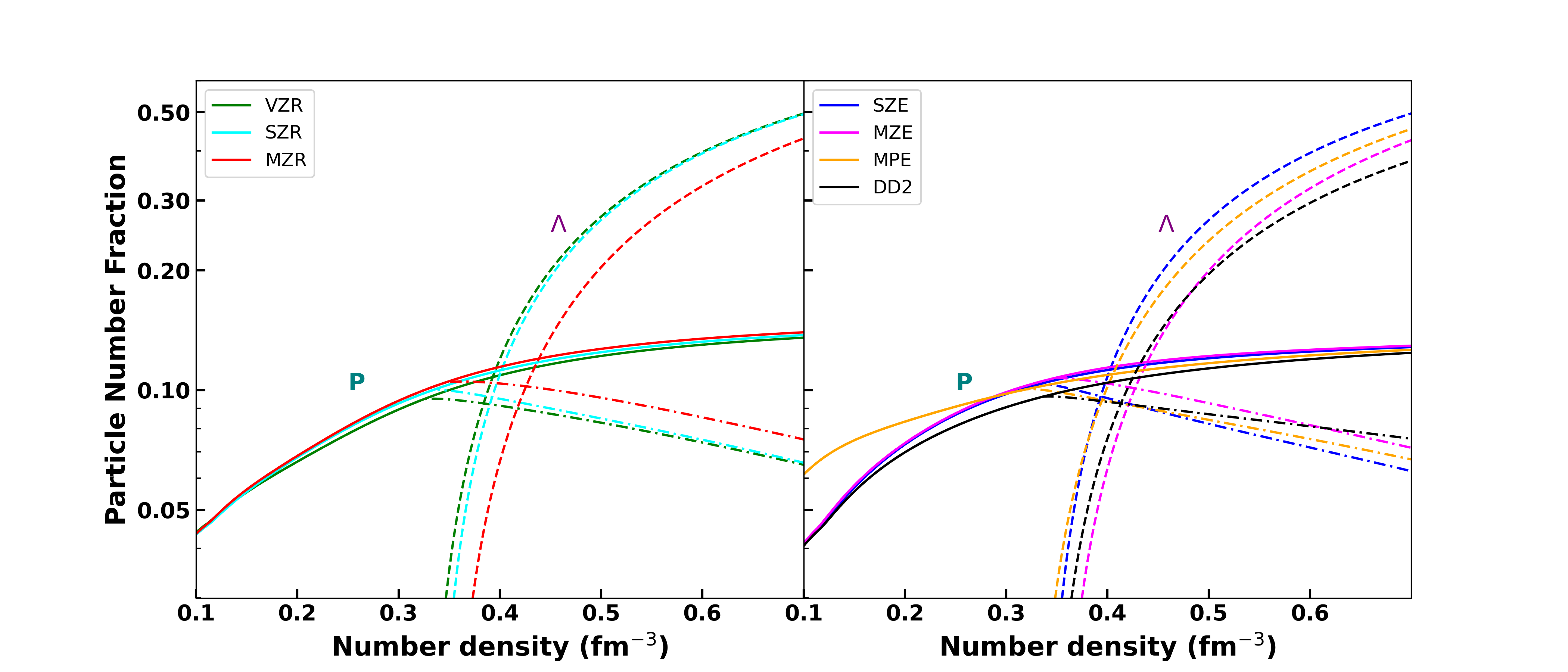}
     \caption{Particle number fractions as a function of baryon number density for various RMF models. Dashed curve shows the  $\Lambda$ hyperon fraction and dashed dot curve shows proton fraction in hyperonic matter (N$\Lambda$), solid curves represent the proton fraction in purely nucleonic matter (N). The left panel displays results for the VZR, SZR, and MZR models, while the right panel shows the SZE, MZE, and MPE models with black lines representing the results for DD2 model.}
     \label{fig:pfrac}
\end{figure*}

In Fig. \ref{fig:pfrac}, we show the proton and the hyperon fractions corresponding to the EOSs presented in  Fig. \ref{fig:EOS}. The proton fractions (solid lines) for nucleonic cases are very similar across different models. While for the hyperonic cases, the proton fraction (dashed dot curves) drops at the onset of $\Lambda$ hyperons (dashed lines).  The onset densities for $\Lambda$'s in each model are listed in Table~\ref{tab:onset}.  We see that the $\Lambda$s appear around the similar density of $\sim 0.34$ fm$^{-3}$ for the scalar cases. While it appears at a slightly higher density of $\sim 0.35$ fm$^{-3}$ for MZE and MZR,  it appears at a lower density of $0.326$ fm$^{-3}$ for MPE. For the VZR case, $\Lambda$s appear at $0.328$ fm$^{-3}$, compared to $0.334$ fm$^{-3}$ for DD2 model. In the left panel, we see that MZR produces fewer hyperons after their relatively late onset resulting in a higher proton fraction than in the other cases.  In the right panel, we find the MPE has higher proton fraction than the others at low densities. But, the earlier appearance of $\Lambda$'s leads to eventual convergence of the proton fractions with the other models. We also draw the proton and $\Lambda$ hyperon fractions for the DD2~\cite{Typel:2009sy} EOS. In comparison, the DD2 exhibits slightly lower proton and $\Lambda$ fractions compared to others. 

\begin{figure*}
     \includegraphics[scale = 0.50]{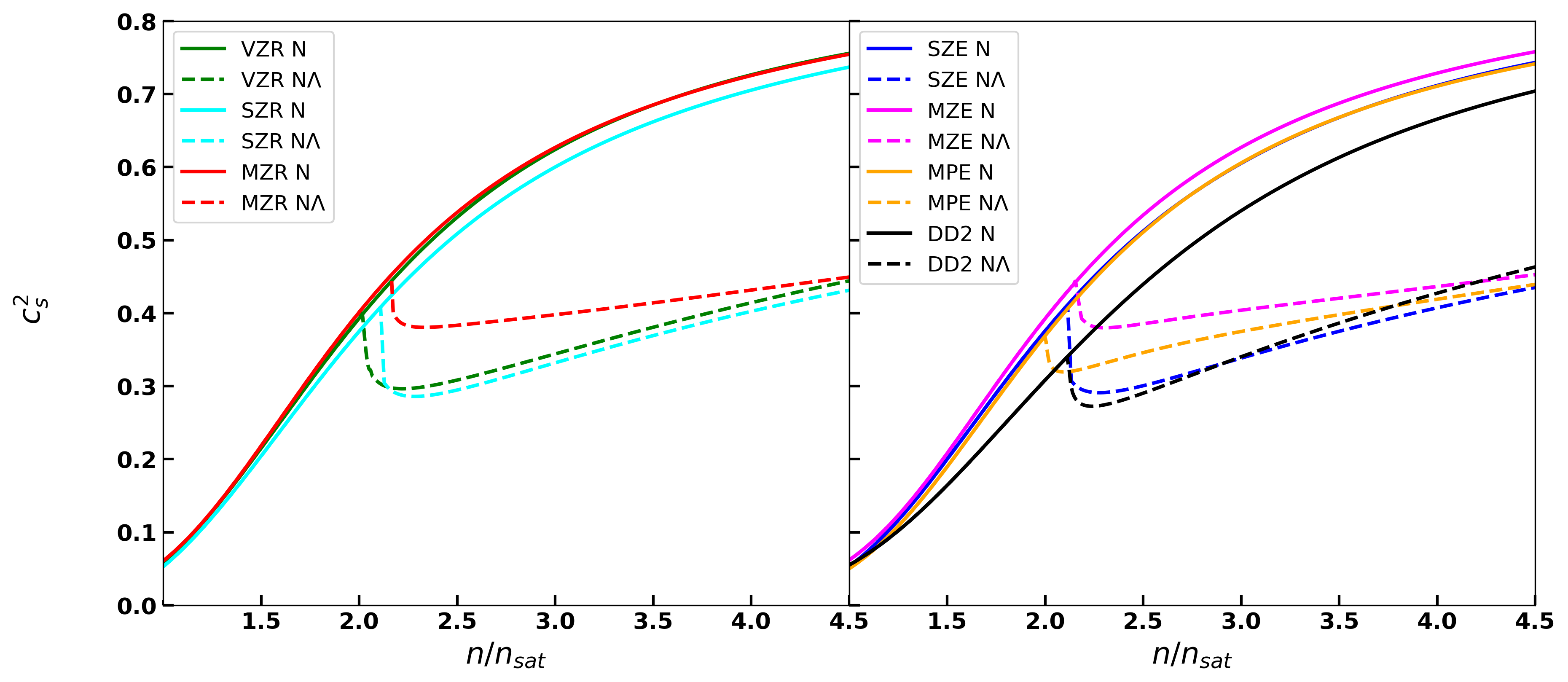}
     \caption{Speed of sound $c_s^2$ as a function of normalized baryon density $n/n_{\rm sat}$. 
    The left panel shows the results for the VZR, SZR, and MZR models, while the right panel presents the SZE, MZE, MPE, and DD2 models. Dashed lines correspond to nucleonic--hyperonic (N$\Lambda$) matter, while solid lines correspond to nucleonic (N) matter.}
    \label{fig:cs2}
\end{figure*}

Furthermore, we calculate the speed of sound for different models and results are displayed in Fig. \ref{fig:cs2}. For nucleonic EOSs, the squared speed of sound ($c_s^2$) increases steadily and attains almost constant values at higher densities. The stiffness controls the speed of sound as expected \cite{Reddy_2018}. The softer EOSs stabilize to lower values of $c_s^2$ at higher densities. The DD2-N EOS has the lowest $c_s^2$ values among them. For their hyperonic counterparts, we see a sudden fall of the $c_s^2$ values at the onset of $\Lambda$s. The decrease in the speed of sound at the threshold can be associated with the stiffness of the hyperon EOSs. SZR and SZE produce the softest hyperon EOSs among the new parameter sets studied in this article. In Fig. \ref{fig:cs2}, these two show the biggest fall of $c_s^2$ at the onset point. The DD2-N$\Lambda$ EOS achieves the lowest $c_s^2$ after the appearance of hyperons.

\begin{figure*}
    \centering
     \includegraphics[scale = 0.50]{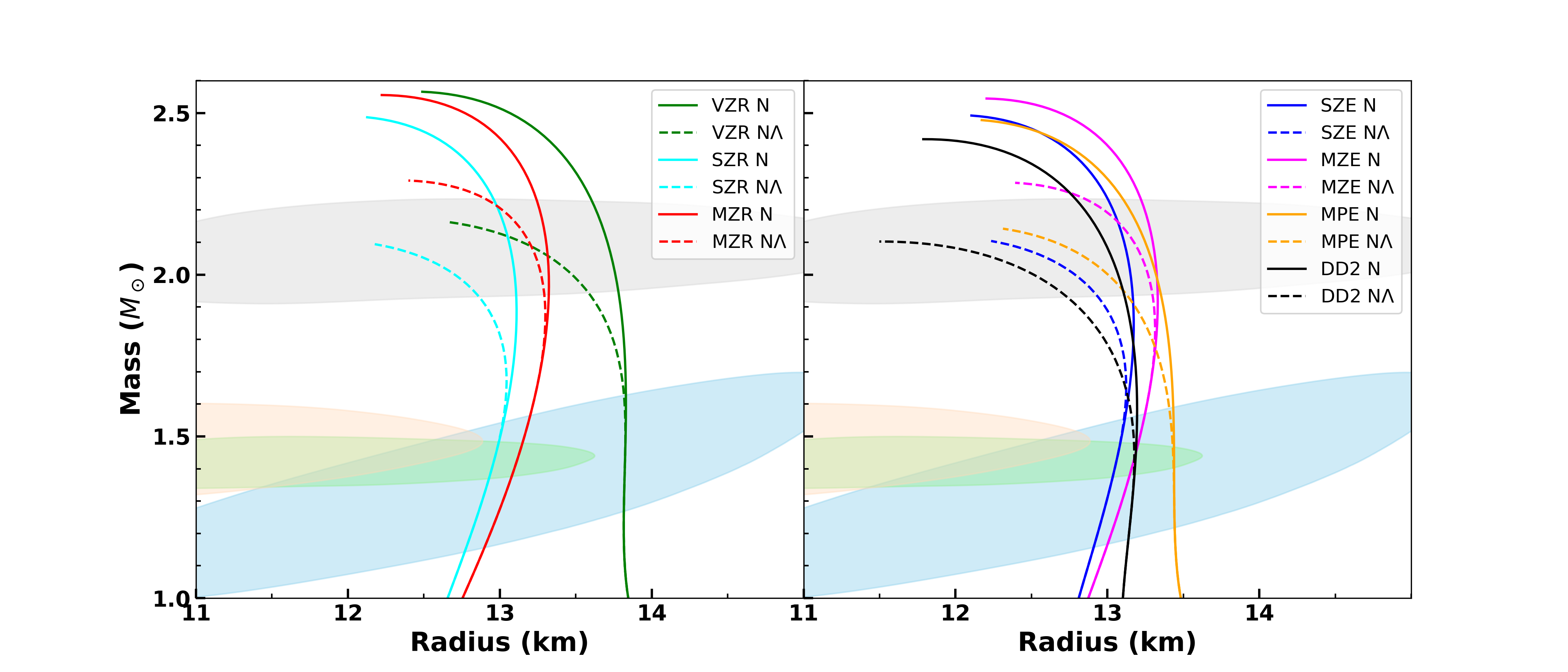}
     \caption{Mass–Radius profiles for various RMF models with observational constraints from NICER. 
     The shaded regions denote the 95\% CI: J0030+0451 (blue), J0437–4715 (green), 
     J0614–3329 (peach), and J0740+6620 (grey). 
     The colour scheme for the EOS model curves is the same as in the previous figures.}
     \label{fig:mr}
\end{figure*}

\begin{figure*}
     \includegraphics[scale = 0.50]{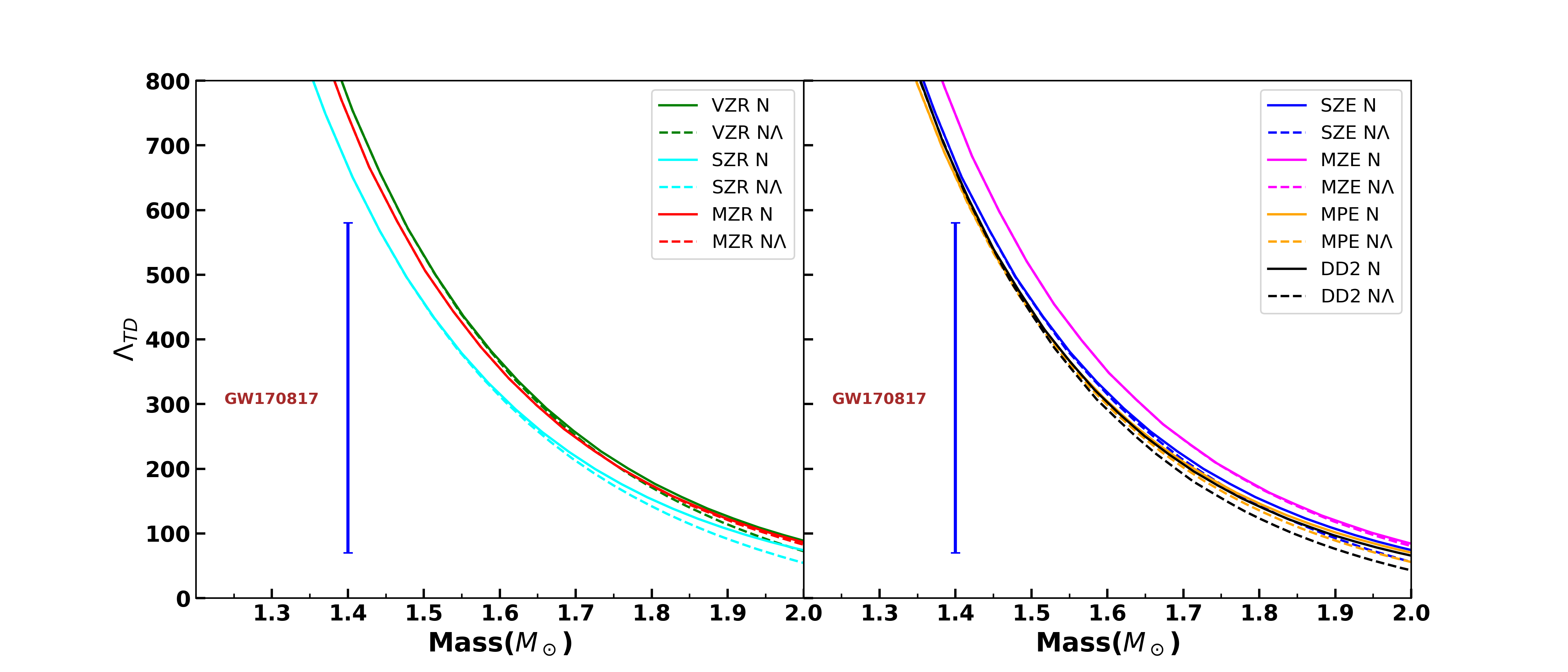}
      \caption{Dimensionless tidal deformability as a function of total gravitational mass for N (solid lines) and N$\Lambda$  (dashed lines). The canonical tidal deformability from GW170817 is shown by vertical blue line.}
     \label{fig:tide}
\end{figure*}

\begin{table*}
 \begin{ruledtabular}
 \begin{tabular}{c|ccccccc}
 \textrm{Models(N)}&  \textrm{SZE}&  \textrm{SZR}&  \textrm{MZE}&  \textrm{MPE}&  \textrm{MZR} &  \textrm{VZR} &  \textrm{DD2}\\
 \colrule
 $n_{onset}$ (fm$^{-3}$) & 0.343 & 0.342 & 0.350 & 0.326 & 0.349 & 0.328 & 0.334\\
 $\epsilon$(MeV fm$^{-3}$) & 349.54 & 347.86 & 356.42 & 328.20 & 355.42 & 324.50 & 329.96\\
 P(MeV fm$^{-3}$)& 46.95 & 46.48 & 53.88 & 36.65 & 54.68 & 41.82 & 36.81\\
\end{tabular}
 \end{ruledtabular}
 \caption{Onset Density of $\Lambda$ hyperons and the corresponding energy density and pressure values at the onset point for different models.}
 \label{tab:onset}
\end{table*}

\begin{table*}
\centering
\begin{ruledtabular}
\begin{tabular}{l *{14}{c}} 
{Models} 
& \multicolumn{2}{c}{SZE} 
& \multicolumn{2}{c}{SZR} 
& \multicolumn{2}{c}{MZE} 
& \multicolumn{2}{c}{MPE} 
& \multicolumn{2}{c}{MZR} 
& \multicolumn{2}{c}{VZR} 
& \multicolumn{2}{c}{DD2} \\
& N & N$\Lambda$ 
& N & N$\Lambda$ 
& N & N$\Lambda$ 
& N & N$\Lambda$  
& N & N$\Lambda$ 
& N & N$\Lambda$ 
& N & N$\Lambda$ \\
\hline

$M_{\rm max}(M_\odot)$ & 2.49  & 2.10  & 2.49 &   2.09  & 2.54 & 2.29  & 2.48  & 2.14  & 2.56  & 2.29  & 2.57  &  2.16  & 2.42 & 2.10 \\
$n_{c_{max}}$(fm$^{-3}$)  & 0.751 & 0.739 & 0.739 & 0.744 & 0.749 & 0.748 & 0.751 & 0.750 & 0.748 & 0.750 & 0.750 & 0.738 & 0.872 & 0.976\\
$R_{M_{max}}$(km) & 12.10 & 12.23 & 12.13 & 12.18 & 12.20 & 12.39 & 12.17 & 12.31 & 12.22 & 12.40 & 12.49 & 12.67 & 11.90 & 11.58\\
\end{tabular}
\end{ruledtabular}
\caption{Maximum mass and corresponding central number density and radius for nucleonic (N) and hyperonic (N$\Lambda$) EOSs, corresponding to each model.}
\label{tab:ns_maxm}
\end{table*}

\begin{table*}
 \begin{ruledtabular}
 \begin{tabular}{c|ccccccc}
 \textrm{Models (N)}&  \textrm{SZE}&  \textrm{SZR}&  \textrm{MZE}&  \textrm{MPE}&  \textrm{MZR} &  \textrm{VZR} &  \textrm{DD2}\\
 \colrule
$n_{c_{1.4 M_\odot}}$(fm$^{-3}$) & 0.334 & 0.335 & 0.324 &  0.340(0.345) & 0.321  & 0.323  & 0.362(0.370)\\
$R_{1.4 M_\odot}$(km) & 13.05  & 12.95 & 13.16 &  13.44 & 13.10 & 13.82 & 13.20 \\
$\Lambda_{1.4 M_\odot}$  & 669.82 & 658.05 & 740.48 & 650.12 & 744.65 & 772.36 & 660.74\\
\end{tabular}
 \end{ruledtabular}
 \caption{The number density, radius and value of tidal deformability are listed for a $1.4 M_{\odot}$ for nucleon (N)-only NS. The values in parentheses are for NS with $\Lambda$ hyperons, if any.}
 \label{tab:1.4}
\end{table*}

Next, we turn our attention to the construction of the stellar configurations using the EOSs shown in Fig. \ref{fig:EOS}. We solve the TOV equations \ref{Eq:TOV} to find the mass-radius sequences for spherically symmetric stars and the tidal deformabilities corresponding to stars in binaries. In Figs. \ref{fig:mr} and \ref{fig:tide}, we have shown the mass-radius and mass-tidal deformabilities sequences, respectively.  

In Fig. \ref{fig:mr}, we show the mass–radius (M–R) profiles  for different RMF models, for both nucleonic (solid lines) and hyperonic (dashed lines) EOSs. In the left panel, among the nucleonic cases, the VZR EOS is the stiffest at high densities, while MZR and VZR behave similarly at lower densities, and SZR remains the softest of the three. This behavior is reflected in the corresponding maximum masses, which are highest for VZR, followed by MZR and SZR, with the stellar radii decreasing in the same order. The inclusion of $\Lambda$ hyperons systematically softens the EOS, resulting in a noticeable reduction in the maximum mass and a slight increase in the corresponding radius compared to the nucleon-only cases. For DD2 model, hyperons lead to a more compact star though. 
Nevertheless, the maximum mass for all three models are above $2 M_\odot$, they are enlisted in Table. \ref{tab:ns_maxm} along with corresponding central number density ($n_{c_{max}}$) and radius ($R_{M_{max}}$). In the right panel, the  MZE parameterization produces the highest mass for the nucleons-only case, followed by SZE and MPE, in accordance with stiffness  pattern seen in Fig. \ref{fig:EOS}. However, this trend between SZE and MPE reverses, with MPE yielding a higher maximum mass due to the earlier onset of hyperons.(See Table. \ref{tab:onset}).

Next, we compare our model results with the latest observations. The 95\% Credible Interval (CI) NICER observations for PSR J0030+0451 (blue), PSR J0437-4715 (green), PSR J0614-3329 (peach) and PSR J0740+6620 (grey) are also displayed in Fig. 4. Overall, the comparison demonstrates that the hyperonic EOSs are capable of simultaneously satisfying the $\sim 2 M_\odot$ mass limit and the NICER  measurements on  PSR J0740+6620, thereby providing viable descriptions of dense matter including hyperonic degrees of freedom. 
The radii predicted by our models at 1.4M$_\odot$ lie between 12.95 km and 13.82 km (see Table~\ref{tab:1.4}), with SZR giving the smallest radius ($12.95$ km) and VZR the largest ($13.82$ km). Note that our $R_{1.4 M_\odot}$ values agree with NICER’s observations of PSR J0437-4715 and PSR J0030+0451 which report $R_{1.4 M_\odot}$ to lie within 11.5–14.5 km at the 95\% credibility level.
The DD2 M–R relations, shown in black, serve as a benchmark for comparison.

The radius of a canonical 1.4M$_\odot$ NS is a key observable for constraining the EOS at densities of $2$–$3 n_{ref}$. 
Its radius, tightly correlated with the tidal deformability measured from GW170817 event and with the mass–radius constraints derived from NICER’s X-ray pulse-profile modeling, both of which provide valuable insights into the EOS of dense matter. From Table \ref{tab:1.4}, we  also observed that the central number densities of 1.4M$_\odot$ ($n_{c_{1.4 M_{\odot}}}$) are less than that of the onset density $n_{onset}$ of $\Lambda$ hyperons for most of the models, except for the mixed density model parameterizations MPE. In DD2 model also, $\Lambda$ hyperons exist in the core of the canonical mass NS. However, their effect is not reflected in the radius or tidal deformability, owing to a negligible fraction of $\Lambda$ hyperons at the relevant central density as evident from Fig. \ref{fig:pfrac}.

In Fig. \ref{fig:tide}, we present the variation of dimensionless tidal deformability as a function of stellar mass for the different RMF models. As expected, the values of the tidal deformability decrease rapidly with increasing mass, reflecting the compactness dependence of the tidal response. Inclusion of $\Lambda$ hyperons slightly reduces the value of tidal deformability at a given mass due to the softening of the EOS. The stiffer models (MZE, MZR, and VZR) predict relatively larger $\Lambda_{1.4 M_\odot}$ values, whereas the softer ones (SZE, SZR, and MPE) fall closer to the central observational range. Our tidal deformabilities values lie between 650.12 and 772.36, which are broadly consistent with constraints from GW170817 \cite{LIGOScientific:2018cki,LIGOScientific:2018hze}.

\section{Summary}
\label{summary}
In this work, we have studied the EOS of NS matter containing hyperons within the RMF formalism with density-dependent meson-baryon couplings. The structure of NS has been calculated for both the purely nucleonic and hyperonic cases. Our main objective of this study is two-fold: first, to study the differences  arising from the density dependence of the couplings  on the scalar density, vector density, and mixed cases, as outlined in section \ref{formalism}. We, then  explore two different functional forms of the $\rho$ meson coupling. We have employed the different parameter sets  for various meson-nucleon couplings. Many of these sets are being used for the first time to study the properties of the NS. We have compared these new results with  those obtained with  the DD2 EOS, which is one of most widely used DDRMF models. Note that, we have further included  $\Lambda$ hyperons in NS matter to compare the qualitative effect of incorporating hyperons within the new density dependencies and the form of the coupling functionals. The incorporation of the full baryon octet should be straightforward and would not change the conclusion of this work.

Most new parameterizations yield stiffer EOSs than DD2 for both nucleonic and hyperonic matter, resulting in stars with larger radii, higher tidal deformabilities, and enhanced $\Lambda$ fractions at high densities. Despite hyperon-induced softening, the maximum masses remain within $2.09$–$2.29 M_\odot$, consistent with NICER observations of J0030+0451 and J0740+6620 at 95\% CI, though marginally inconsistent with the 90\% tidal deformability constraint from GW170817. These findings underscore the significance of exploring alternative density dependencies in meson–baryon couplings and pave the way for future Bayesian analyses, following the prescriptions from Refs. \cite{Char:2023fue,Char:2025zdy,Char:2025nli} to further constrain the parameter space using multi-messenger data.

\section*{Acknowledgments}
 AS and SB acknowledge the Department of Science \& Technology, India for the support  via project no: CRG/2022/008360. AS also thanks Debanjan Guha Roy and Anagh Venneti for their valuable guidance and insightful discussions throughout the course of this work. SG acknowledges the partial support of CSIR vide project no. 03/1513/23/EMR-II. This project  has received funding from the European Union’s Horizon 2020 research and innovation programme under the Marie Skłodowska-Curie grant agreement No. 101034371. PC acknowledges the support from the European Union's HORIZON MSCA-2022-PF-01-01 Programme under Grant Agreement No. 101109652, project ProMatEx-NS.

\bibliography{References}

\end{document}